\documentstyle[twocolumn,aps,epsf,amsmath,amssymb]{revtex}
\begin{document}

\twocolumn[
\hsize\textwidth\columnwidth\hsize\csname@twocolumnfalse\endcsname
\draft

\title{A low density finite temperature apparent ``insulating'' phase
  in 2D systems} 
\author{S. Das Sarma and E. H. Hwang}
\address{Condensed Matter Theory Center, 
Department of Physics, University of Maryland, College Park,
Maryland  20742-4111 } 
\date{\today}
\maketitle

\begin{abstract}
We propose that the observed low density ``insulating'' phase of a 2D
semiconductor system, with the carrier density being just below ($n
< n_c$) the so-called critical density where the derivative of
resistivity changes sign at low temperatures (i.e. resistivity
$\rho(T)$ increases with increasing $T$ for $n > n_c$ whereas it
decreases with increasing $T$ for $n < n_c$), is in fact a
``high-temperature'' crossover version of the same effective metallic 
phase seen at
higher densities ($n>n_c$). This low density ($n<n_c$) finite
temperature crossover 2D effective insulating phase is characterized
by $\rho(T)$ with power law temperature dependence in contrast to the
truly insulating state (occurring at still lower densities) whose
resistivity increases exponentially with decreasing temperature. 


\end{abstract}
\vspace{0.5cm}
]

\newpage

The so-called 2D metallic phase, first identified \cite{Kravchenko94} in n-Si
MOSFETs by Kravchenko and collaborators, is still a subject of
considerable interest and controversy \cite{Abrahams01}. Although the
effectively metallic character (albeit at finite temperatures) of this
phase is well-established experimentally in many 2D systems, it is
still not known whether a true $T=0$ 2D metal exists. The nature of
the associated 2D metal-insulator-transition (2D MIT) as a function of
carrier density ($n$) is also a mystery (particularly since a
noninteracting disordered 2D system at $T=0$ is known on firm
theoretical grounds not to undergo any MIT as all states in 2D are
localized in the presence of finite disorder \cite{Abrahams79}). 
There is increasing
evidence that the putative 2D ``metallic'' phase, while being a perfectly
good ``effective'' metal down to $T=50-100mK$ where experiments have so
far been carried out, is essentially a finite temperature effective
metal with screening \cite{DH1}
and related electron-electron interaction effects \cite{zala}
playing important quantitative roles in determining the temperature
dependent resistivity (and producing the effective metallic
behavior). It should also be emphasized that no reasonable model or
theory for a $T=0$ strongly-correlated 2D metal (stabilized presumably
by electron-electron interaction in the presence of disorder) has been
put forward in the literature\cite{Abrahams01}. 
(Such a true 2D metal, if it exists at
$T=0$, cannot be a Fermi liquid since the corresponding noninteracting
2D system is an insulator at $T=0$ \cite{Abrahams79}.) 
The observed 2D metallic phase
shows strong temperature dependence ($d\rho/dT > 0$) which is
nominally unexpected in a metal at low temperatures where phonons
freeze out in the low temperature Bloch-Gr\"{u}neisen regime. We have
argued, on the basis of concrete calculations, that the strong
temperature dependence in the resistivity in the effective metallic
phase is caused by a temperature dependent effective disorder arising
from the strong temperature dependence of low density 2D
screening \cite{DH1}. 
This view point of the metallic temperature dependence has
received further support from recent theoretical calculations of
interaction corrections to 2D resistivity generalizing the screening
theory \cite{zala}. 
Unlike 3D metals, where $q_{TF}/2k_F \sim 1$ and $T/T_F \ll 1$
(where $q_{TF}$, $k_F$, $T_F$ are the Thomas-Fermi screening wave
vector, the Fermi wave vector, and the Fermi temperature,
respectively), the semiconductor-based 2D systems at low carrier
densities could typically have $q_{TF}/2k_F \gg 1$ and $T/T_F \sim 1$,
engendering strong screening-induced temperature dependence in the
resistivity arising from the strong temperature dependence in the
screened impurity disorder scattering in contrast to 3D metals where
any temperature dependence of disorder scattering is exponentially
suppressed at low temperatures. 

In this paper we extend our screening calculations to very low
densities (below $n_c$, the apparent critical density for the MIT),
where the experimental resistivity increases with decreasing
temperature ($d\rho/dT <0$) indicating the existence of a nominally
insulating phase. Quite surprisingly we predict the existence of an
effective ``insulating'' phase defined by $d\rho/dT <0$ at intermediate
$T$ and {\it low n}, arising from the same temperature dependent
disorder effect which produces the effective metallic phase
($d\rho/dT >0$) at {\it low T} and {\it high n}. We find that for a
range of densities below $n_c$ our calculated resistivity,
$\rho(T,n)$, as a function of temperature ($T$) and density ($n$),
exhibits striking qualitative similarity to
the experimentally observed $\rho(T, n\le n_c)$
\cite{noh02,noh03,lilly,hanein,mills}, 
where the resistivity
increases with decreasing $T$ (i.e. $d\rho/dT <0$) in a rather slowly
rising approximate
power law manner in contrast to the thermal exponential (either
Mott variable range hopping or activated) behavior expected of a true
$T=0$ insulator.
The strongly insulating phase (which is beyond the
scope of our theory) may occur at still lower densities where
$\rho(T)$ would rise 
exponentially with decreasing temperature. Our theory
applies to all 2D systems where the 2D MIT has been reported, and
most significantly, in some systems (such as high mobility 2D $n$- and
$p$-GaAs) the observed 2D insulating phase seems to be entirely the
novel low density effective insulating phase 
identified in this paper as the exponential insulating behavior seems
not to manifest itself in high-quality GaAs systems except perhaps at
the lowest densities and temperatures.
An important salient feature of this low
density novel phase (with $d\rho/dT <0$) is that it is purely a
temperature-induced crossover behavior, and as such the critical
density $n_c$ at which the 2D system crosses over to this phase (as
$n$ is decreased from above $n_c$ to below) is completely
non-universal, being a function of  temperature itself, i.e. $n_c
\equiv n_c(T)$, with the effective critical density $n_c$
progressively decreasing with decreasing carrier temperature. The
experimentally determined ``critical'' density is then $n_c$($T=T_{\rm
  min}$)
where $T_{\rm min}$($\approx 50-300 mK$ depending on the experiment) 
is the effective minimum
temperature to which the 2D carrier system (and {\it not} the
background bath) can be cooled down (typically $T_{\rm min} \gg T_{\rm
  bath}$
in 2D semiconductor systems for low values of  $T_{\rm bath}$). Such a
non-universal behavior for the critical density defining 2D MIT, namely
that $n_c \equiv n_c(T)$ with the observed effective $n_c$ decreasing
with decreasing temperature, has been pointed out 
experimentally \cite{hanein,mills}. In
particular, our theory specifically rules out a sharp (temperature
independent) separatrix delineating the effective metallic ($d\rho/dT
>0$) and the effective insulating ($d\rho/dT <0$) phase in the 2D MIT
phenomenon. We emphasize that the experimental evidence for a sharp
separatrix in the 2D MIT is extremely sparse, and most 2D MIT
experiments observe a temperature dependent crossover density $n_c(T)$
separating $d\rho/dT >0$ for $n>n_c$ and $d\rho/dT <0$ for $n<n_c$
in agreement with our theory.

We refer to this novel (low $n$/intermediate $T$) crossover
``insulating'' ($d\rho/dT <0$) phase as an ``apparent'' insulating phase in
order to emphasize the fact that this phase is $not$ a true insulating
phase at $T=0$, but is rather a low density (and ``high'' temperature)
metallic phase where $\rho(T)$ decreases with increasing temperature
in a power law fashion. This newly identified effective insulating
phase (i.e. $d\rho/dT <0$) is only an apparent phase since lowering
temperature further (at a fixed $n<n_c$) will eventually cause a
temperature-induced re-entrance into the effective metallic phase at
sufficiently low temperatures which, however, may not be accessible
experimentally making the apparent insulating phase to appear to be a
true insulating phase due to this low-$T$ cut off. We emphasize that
the hallmark of our proposed apparent insulating phase is 
the approximate power
law temperature dependence of the resistivity in contrast to a true
insulating phase where $\rho(T)$ diverges exponentially with
decreasing temperature. 
All localization effects, either weak or strong, are neglected in our
theory. This is entirely consistent with our motivation of trying to
understand carrier transport in a ``high temperature phase'' where
quantum interference effects should be small. The Boltzmann theory,
described below, should be well-valid for this effective
high-temperature semiclassical phase independent of the actual value
of the resistivity.

We calculate the resistivity of a 2D carrier system scattering off
random charged impurity centers with a screened Coulomb impurity
scattering potential. Within the Drude-Boltzmann semiclassical
transport theory with the carrier-impurity interaction treated in the
standard ensemble averaged Born approximation, we can express the 2D
resistivity $\rho$ as \cite{ando}
\begin{equation}
\rho^{-1} = n e^2 \langle \tau \rangle/m,
\end{equation}
where $m$ is the carrier effective mass assuming a parabolic band
effective mass approximation for the semiconductor band structure with
$n$ being the 2D carrier density. The thermally averaged scattering
time $\langle \tau \rangle$ can be written, within the Drude-Boltzmann
semiclassical theory (and the standard relaxation time
approximation), as 
\begin{equation}
\langle \tau \rangle \equiv \langle \tau \rangle_{E} = \frac{\int dE
  \tau(E) E \left ( -\frac{\partial f}{\partial E} \right )}{\int dE E \left
    (-\frac{\partial f}{\partial E} \right )},
\end{equation}
where $\langle \cdot \cdot \cdot \rangle_E$ indicates an energy
averaging over the carrier thermal distribution function $f(E) =
[1+\exp\{\beta(E-\mu)\}]^{-1}$ with $\mu \equiv \mu(T,n)$ as the
finite temperature chemical potential ($\beta = 1/k_BT$) given by $\mu =
\frac{1}{\beta}\ln[-1 + \exp(\beta E_F)]$ for a 2D system (note that
$\mu=E_F$ at $T=0$, but $\mu$ at finite T could be substantially
different and is negative for $T>1.4T_F$). The central quantity of
interest is therefore the energy dependent transport relaxation time
(i.e. the scattering time) $\tau(E)$, which in the Born approximation
is given by 
\begin{eqnarray}
\tau^{-1}(E)|_{E=\epsilon_{\bf k}} &=& \frac{2\pi}{\hbar}\sum_{{\bf
    k}'}\sum_{\alpha}\int_{-\infty}^{\infty}dz N_i^{(\alpha)}(z)
|u^{(\alpha)}_{{\bf k}-{\bf k}'}(z)|^2  \nonumber \\
&\times& (1-\cos\theta_{{\bf k k}'})
\delta(\epsilon_{\bf k}-\epsilon_{{\bf k}'}),
\end{eqnarray} 
where $\epsilon_{\bf k}=\hbar^2k^2/2m$ is the effective mass band
energy of the 2D carriers with $k\equiv |{\bf k}|$ as the 2D wave
vector, and $N_i$ is the three dimensional volume density of the
random charged impurity centers of the $\alpha$-th kind (we allow, in
general, of different kinds of quenched disorder in the problem
arising from the distributions of different types of random charged
impurity centers in the 2D system) --- we take $z$ direction to be the
confinement direction perpendicular to the relevant 2D plane of
confinement of the carrier system. In Eq. (3) $u_q^{(\alpha)}(z)$ is
the 2D (in the $x$-$y$ plane) Fourier transform of the 3D screened
electron-impurity interaction in the system defined to be 
\begin{equation}
u_q^{(\alpha)}(z_i) = \frac{2\pi Z^{(\alpha)}e^2F(q,z_i)}{\bar\kappa
  q\epsilon(q)},
\end{equation}
where the effective dielectric function $\epsilon(q)$ is given by
\begin{equation}
\epsilon(q) = 1 - \frac{2\pi e^2}{\bar \kappa q} f(q) \Pi(q),
\end{equation}
where $\Pi(q)$ is the finite temperature static 2D noninteracting
polarizability function and $\bar \kappa$ as the background static
lattice dielectric constant. 
(We have carried out calculations also by
including several different local field corrections in the 2D
polarizability, but our qualitative results remain unaffected and
therefore we will only present here RPA screening results where $\Pi$
is the 2D non-interacting static polarizability.) In Eqs. (4) and (5)
$F(q,z_i)$ and $f(q)$ are respectively the 2D subband form-factors for
electron-impurity and electron-electron interactions, associated with
the finite width of carrier wavefunctions in the $z$ direction. These
form factors are essential for quantitative accuracy, most
particularly at the very low carrier densities of our interest where
the quasi-2D electron layer spreads considerably in the transverse
direction making the strict 2D approximation highly unreliable. We
calculate the form factors by using the Fang-Howard-Stern variational
wavefunctions for the quasi-2D layer \cite{ando}. 
For more details on the theory see refs. \onlinecite{ando} and
\onlinecite{DH03}.

We have carried out extensive calculations of $dc$ transport, based on
the above-described Boltzmann theory, in several different 2D systems
of current interest (i.e. systems where an apparent MIT, as reflected
in the change of sign in $d\rho/dT$ at some low density, has been
experimentally observed) at low temperatures ($T=50mK-  5K$) and
densities ($n$ around, $n\sim n_c$, the reported critical densities
being $n_c\approx 10^{11},$ $10^{10}$, $10^9 cm^{-2}$ for typical high
quality n-Si-MOS, p-GaAs, and n-GaAs systems respectively). Before
presenting our numerical results for $\rho(T,n)$ in Figs. 1 - 3, we
first discuss qualitatively the theoretical results expected on the
basis of Eqs. (1) - (5) in various temperature and density regimes. The
dimensionless temperature, $T/T_F \sim n^{-1}$, plays an important role
in our calculated temperature and density dependence 2D resistivity
$\rho(T,n)$.

First, a straightforward asymptotic expansion of the Boltzmann
transport equation shows that, for $T/T_F \ll 1$, the resistivity can
be written as $\rho = \rho_0 + \Delta \rho(T)$, where
$\rho_0=\rho(T=0)$ and $\Delta \rho$ is given at very low temperature
by  \cite{DH03,SD}
\begin{equation}
\frac{\Delta \rho(T \ll T_F)}{\rho_0} \approx A_1 \left (
  \frac{T}{T_F}\right ) + A_{3/2} \left ( \frac{T}{T_F} \right )^{3/2}
- B_2 \left( \frac{T}{T_F} \right )^2, 
\end{equation}
where $A_1(>0)$, $A_{3/2}(>0)$ arise from the thermal smearing of the
electronic polarizability at $2k_F$ wave vector, and $B_2 (>0)$ arises
from the thermal energy averaging over the Fermi surface. Note that
the first two screening terms in Eq. (6) are both positive, indicating
an increasing resistivity with increasing temperature (the so-called
2D metallic behavior) due to the suppression of screening at finite
temperatures whereas the third term, arising from the thermal energy
averaging in Eq. (2), is a negative correction to resistivity (i.e. a
positive correction to conductivity) due to the increased thermal
velocity of the carriers at finite temperatures. The expansions for
$A_1$, $A_{3/2}$, and $B_2$ have earlier been derived in the
literature \cite{zala,DH03,SD}: 
\begin{equation}
A_1 = 2 \left ( 1+ \frac{1}{fq_0} \right )^{-1},
\end{equation}
\begin{equation}
A_{3/2} = 2.646 \left ( 1+\frac{1}{fq_0} \right )^{-2},
\end{equation}
\begin{equation}
B_2 = \frac{\pi^2}{6} p (p+1),
\end{equation}
where $q_0 = q_{TF}/2k_F$, $f \equiv f(q=2k_F)$ is the 2D subband
form-factor (defining the deviation from the strict two dimensionality
in the layer) at wave vector $2k_F$, and $p$ $(\sim 1)$ gives the energy
dependence of the scattering time in Eq. (3), i.e. $\tau(E) \sim E^p$
for $E \sim E_F$. 
Three important feature of the asymptotic expansion given in Eq. (6)
need to be emphasized: (1) The $A$ and the $B$ terms in Eq. (6) arise
from totally different physics, and therefore we have neglected a
possible $A_2(T/T_F)^2$ term in the screening expansion in Eq. (6)
which may very well modify the $B_2$ term to $B_2' \equiv
(B_2-A_2)$. We use Eq. (6) only for the purpose of qualitative
discussion, and all our results are based on the full numerical
calculation of the semiclassical transport equations (Eqs. (1) - (5)) making
any qualitative difference between  $B_2$ and $B_2'$ irrelevant for our
results. (2) There are higher-order interaction corrections for all
the terms in Eq. (6), but these corrections are quantitatively unknown
and are therefore left out of our discussion completely. (3) The
asymptotic expansion only holds at very low temperature ($T/T_F \ll
1$) which may not be of relevance to the 2D MIT phenomena \cite{DH03}.  
Details on the low temperature transport behavior can be found in our
recent work \cite{DH03}. Here we are interested in the ``high
temperature'' resistivity, which is manifestly not given by Eq. (6).

It is possible to obtain the asymptotic high temperature result for
the classical limit ($T \gg T_F$) of the 2D resistivity defined
through Eqs. (1) - (6), and the result is  \cite{DH1}
\begin{equation}
\rho(T\gg T_F) \sim (T_F/T),
\end{equation}
indicating that the ``high temperature'' resistivity decreases with
increasing temperature in an apparently ``insulating-like'' manner
(i.e. $d\rho/dT <0$ for $T \gg T_F$). This can be rather easily
understood physically based on the fact that at high temperatures the
main effect arises from the thermal increase in the average carrier
velocity which obviously leads to a decreasing resistivity. 
Thus, the linear power law resistivity, $\rho \sim T^{-1}$, in
Eq. (10) arises from the energy averaging in Eq. (2).

Even before carrying out the full numerical 
calculations it is obvious that any smooth interpolation
between the ``low
temperature'' behavior of Eq. (6) and the ``high temperature'' behavior of
Eq. (10) would exhibit non-monotonic temperature dependent 
resistivity \cite{DH1}
associated with this quantum-classical crossover, which has been seen
\cite{noh02,noh03,lilly,hanein,mills} experimentally.
This leads to the following generic temperature dependence of
$\rho(T/T_F)$: At very low temperatures $\rho$ is linear in $T/T_F$
with $d\rho/dT$ ($>0$) being determined by the absolute value of the
coefficient $A_1$; with increasing $T/T_F$ (but with $T/T_F <1$) it is
possible (but {\it not} necessarily essential) for $d\rho/dT$ to
change sign becoming negative, $d\rho/dT<0$, if the $B_2$ term can
overcome the effects of the first two positive terms within the
constraint $T/T_F <1$. This obviously depends on the relative values
of coefficients $A_1$, $A_{3/2}$, and $B_2$. In Si (100) inversion
layer and in p-GaAs system screening effects are strong leading to
$A_1$, $A_{3/2}\gg B_2$ so that at low $T/T_F$ the metallic
resistivity, $d\rho/dT >0$, dominates. In n-GaAs 2D system, on the
other hand, screening effects are weak (since the density of states is
small) and for modest values of $T/T_F$, the $B_2$ term is
larger than the first two terms, particularly at high densities where
$2k_F \gg q_{TF}$, leading to $d\rho/dT <0$ for moderate
values of $T/T_F$. (In fact, this is also expected in highly
disordered systems where screening effects may be cut off 
by impurity scattering at a
characteristic temperature $T_D$ which may not be too low.) The
message from the analytical results of Eqs. (6) - (9)  is therefore
clear: A 2D system would always exhibit ``metallic'' resistivity (defined
as $d\rho/dT >0$) for the lowest value of the dimensionless
temperature $T/T_F$ (ignoring all weak localization effects which we
neglect throughout this paper) crossing over to an apparent ``insulating''
(defined as $d\rho/dT <0$, but {\it not} with an exponential
temperature dependence) resistivity at some intermediate temperature,
$T_{\rm cross}$, where ``metallic'' ($d\rho/dT >0$) screening effects are
overwhelmed by the ``insulating'' ($d\rho/dT <0$) thermal averaging
effects. If the carrier density of the 2D system is effectively low so
that $T_{\rm cross}$ is lower than the lowest achievable carrier
temperature ($T_{\rm min}$) in the system, then the 2D system will
manifest an apparent insulating phase ($d\rho/dT <0$) down to the
lowest experimental temperature $T$ ($>T_{\rm cross}$) since the low
temperatures ($T<T_{\rm cross}$) where the screening effect dominates for
a crossover to the metallic behavior ($d\rho/dT >0$) cannot be
achieved. We note that it is possible in many situations, depending on
carrier densities and system parameters (e.g. effective mass, type of
impurity scattering, etc.), for $T_{\rm cross} < T_F$. Remembering that
$T_F \approx 0.7 \tilde n$ K (Si and p-GaAs) and $4 \tilde n$ K (n-GaAs)
where $\tilde n$ is 
the 2D carrier density measured in the units of $10^{10} cm^{-2}$, we
conclude that $T_{\rm cross}$ could be as low as $100mK$ (in some
situations) for 2D carrier densities around $10^{10}cm^{-2}$ for n-Si
(and p-GaAs) and $10^9 cm^{-2}$ for n-GaAs systems. A characteristic
fundamental feature for the crossover we are discussing here is the
strong dependence of the crossover temperature on the carrier density
with $T_{\rm cross}(n)$ increasing strongly with carrier density. The
other important feature is the 
non-existence of the apparent
insulating ($d\rho/dT <0$) phase at low enough temperature
($T<T_{\rm cross}$), i.e. if the carrier 
temperature of a 2D system can be
decreased arbitrarily then (leaving out all localization effects) the
system would always become metallic ($d\rho/dT >0$) at low enough
temperatures (defined as $T \ll T_{\rm cross}$).

Our numerical transport results shown in Figs. 1 - 3 for 
p-GaAs and n-GaAs respectively bear out the theoretical
considerations outlined above. In particular, the calculated
resistivity is always metallic at the lowest temperatures, manifesting
a clear crossover to an insulating behavior at $T_{\rm cross}(n)$ with a
change of sign in $d\rho/dT$ (i.e. $d\rho/dT >0$ for $T<T_{\rm cross}$ and
$d\rho/dT<0$ for $T>T_{\rm cross}$) as described above. 
We note that $T_{\rm cross}$
could be quite

\begin{figure}
\epsfysize=2.5in
\centerline{\epsffile{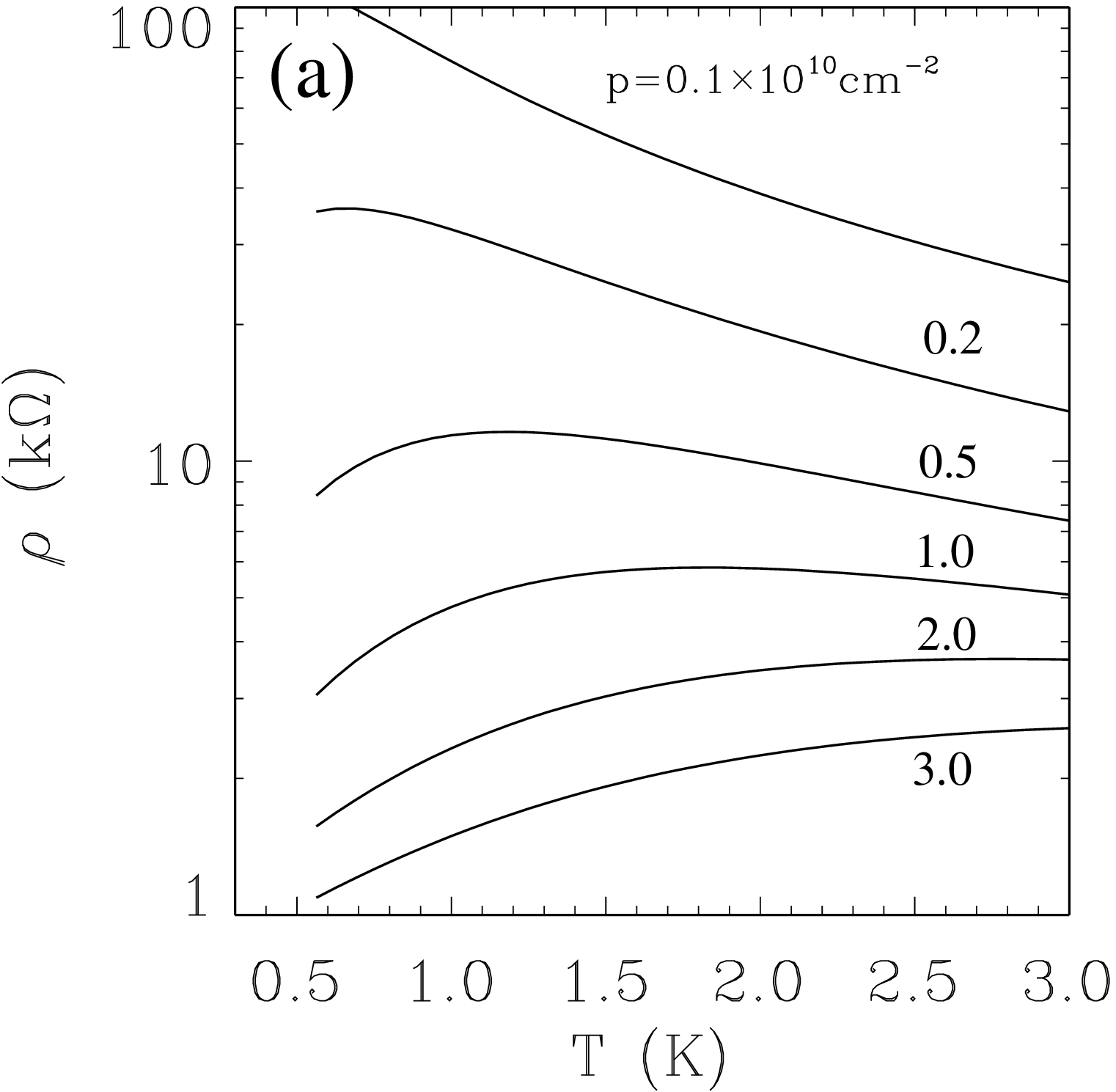}}
\epsfysize=2.5in
\centerline{\epsffile{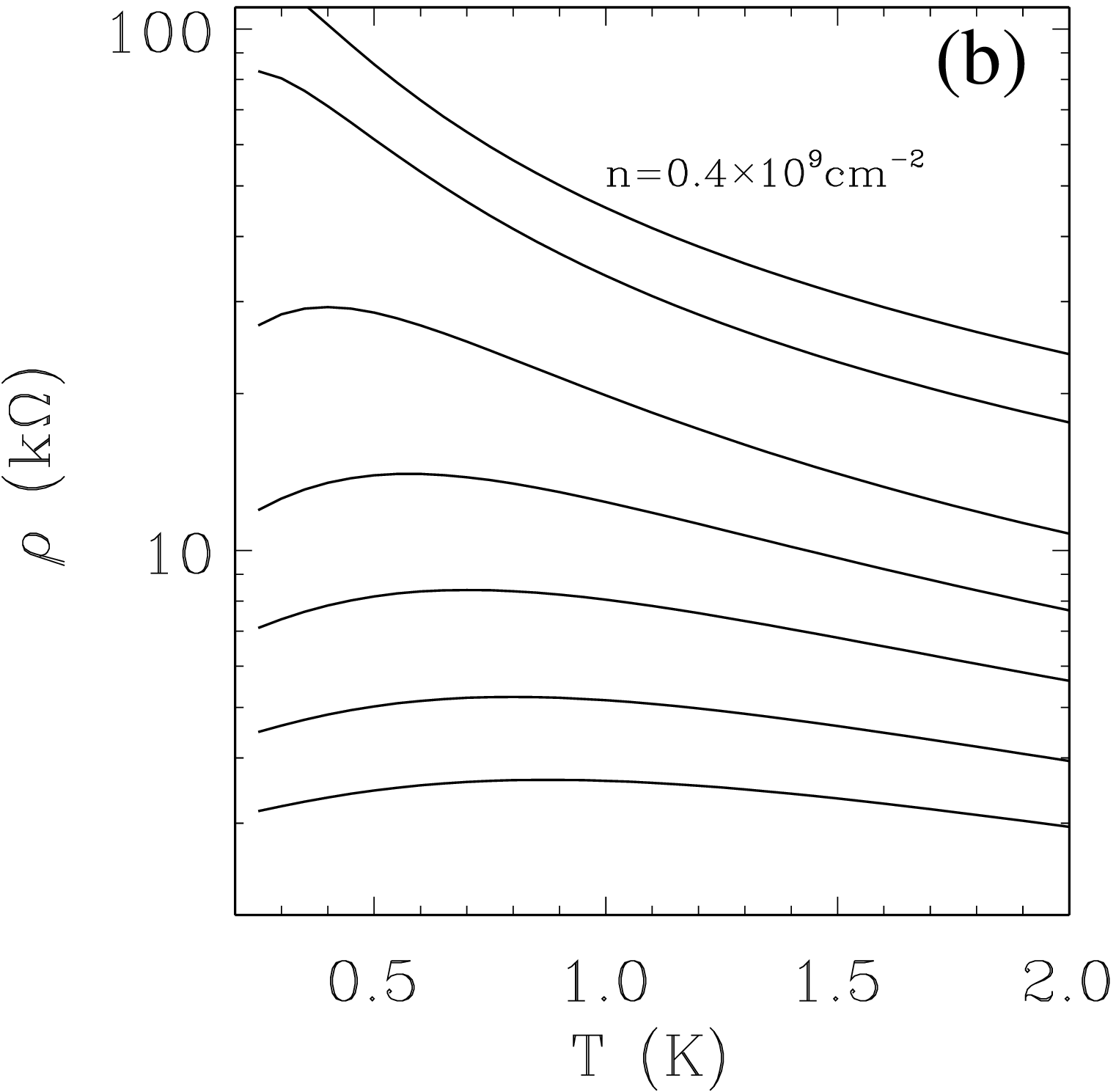}}
\caption{
(a) Calculated 2D resistivities for p-GaAs system for various hole
densities $p=$0.1, 0.2, 0.5, 1.0, 2.0, 3.0$\times 10^{10}cm^{-2}$
(from the top) as a function of temperature.
In this figure we use the charged bulk impurities inside the quasi-2D
systems.
(b) Calculated 2D resistivities for n-GaAs system for various electron
densities $n=$0.4, 0.5, 1.0, 2.0, 3.0, 4.0, 5.0$\times 10^{9}cm^{-2}$
(from the top) with the charged interface impurities
as a function of temperature. ($T_D=0$ is used for these results.)}
\label{fig1}
\end{figure}

\noindent
low for low values of carrier density, and therefore it
is possible, in principle, for the apparent insulating behavior to
arise entirely from this crossover effect. In particular, in
Fig. 1 we choose (rather arbitrarily) the lowest possible
achievable carrier temperature to be $300mK-500mK$ (for the purpose of
illustration) and show what a typical
2D MIT transport data set (in the $T=300 mK -  3K$ range) would look
like based entirely on this finite temperature crossover
phenomenon. The overall qualitative consistency between our results
and many of the existing 2D MIT experimental
results in the literature suggests that the simple (and physically
appealing) crossover scenario presented 
in this  paper cannot be
entirely ruled out as the cause for the 2D MIT behavior observed
experimentally at least in some situations. 
We note a quantitative discrepancy between our theory and experiment:
in general, we find $T_{\rm cross}$ in our theory to be higher than
the corresponding crossover temperature in the experiment. (This
discrepancy can be corrected by introducing a cut-off $T_D$ in the
screening as discussed later in this paper.)

It is essential to emphasize two
points: (1) The effective insulating phase is only an apparent
insulating phase by virtue of the insulator being defined through the
negative temperature derivative of resistivity, i.e. $d\rho/dT
<0$. This apparent insulating phase is really a metallic
(i.e. extended or delocalized electron wavefunction) electron liquid
phase (interacting with random charged impurity centers) exhibiting
$d\rho/dT <0$ down to a crossover temperature scale ($T>T_{\rm cross}$)
with the lowest temperature ($T \ll T_{\rm cross}$) phase being always a
true metal with $d\rho/dT >0$. (2) The temperature dependence of
$\rho(T) $ in this apparently insulating phase 
simulates a power law (with $\rho \sim 1/T$ in the $T>T_F$
high-temperature region) rather
than the exponential temperature dependence typical of a true
insulating phase. 
The finite temperature crossover ``insulating''
phase (i.e. $d\rho/dT <0$) is, by definition, a ``high'' temperature phase
in our Drude-Boltzmann transport model with the true ``low'' temperature
($T\rightarrow 0$) phase in our theory (which manifestly excludes weak
and strong localization effects) being always a metallic
(i.e. $d\rho/dT >0$) phase. The crucial point of physics is, however,
the fact that experimentally there is always a low temperature
cut-off, $T_{\rm min}$, below which carriers simply cannot be cooled
down, and for $T_{\rm cross} < T_{\rm min}$, the 2D apparent ``insulating''
phase discussed in this paper would, for all particular purposes,
behave like a 'real' insulating phase. The ``apparent'' 2D metal-insulator
transition in this crossover scenario will therefore appear to occur
at the density $n_c$ defined by $T_{\rm cross}(n_c) \approx
T_{\rm min}$.

In comparing our theoretical results with the existing 2D MIT data in
the experimental literature, we mention that a large number of
reported data look qualitatively similar to the results shown in
Figs. 1 - 3 of this paper. In particular, the following features of our
theoretical results are often observed experimentally: (1) the
low-density-low temperature 2D insulating phase (for $n<n_c$) is often
seen to have an approximate
power law temperature dependence resistivity; (2) the
low temperature extrapolated ($T\rightarrow 0$) resistivity does not
seem to be divergent in many cases; (3) the experimentally
observed critical density $n_c$ has been reported to be a function of
the lowest measurement temperature with $n_c(T)$ decreasing
monotonically with decreasing temperature.
For low enough carrier densities, most 2D systems 
should eventually enter a strongly
localized transport regime where the low-temperature resistivity
diverges in an exponential manner, but in high quality (i.e. low
disorder) systems of interest in the 2D MIT phenomenon such
exponentially divergent resistivity seems to be always preceded by a
density regime (below $n_c$) where $\rho(T)$ increases smoothly in a
power law manner with decreasing temperature. This intermediate
density regime of apparent insulating behavior is most strongly
manifested in 2D p-GaAs and n-GaAs systems, but it has also been
seen in n-Si MOS systems. A possible reason for the prevalence of the
apparent insulating phase in 2D GaAs systems could be the much higher
quality (i.e. much lower disorder) for GaAs-based 2D systems compared
with 2D Si structure, making it possible for the system not to
manifest localization down to rather low carrier densities,
thus enabling the crossover insulating phase to show up more easily
without being preempted by localization.

In discussing experimental data, it is important to realize that the
measured 2D resistivity $\rho(T)$ typically shows a saturation at low
enough temperature, i.e. below some saturation temperature ($T_{s}\sim
50-300 mK$, depending on the experiment) 
$\rho(T<T_s)$ becomes essentially a constant completely
independent of temperature. Although an underlying (unknown)
fundamental cause for this resistivity saturation behavior cannot be
definitively ruled out, the saturation is generally believed to arise
from a saturation in the carrier temperature due to the inevitable
carrier heating problems in semiconductors at low temperatures and
densities. This possibility of carrier temperature saturation at low
temperatures (where further cryogenic cooling only lowers the
temperature of the surrounding bath, not the 2D electrons themselves)
is further supported by the fact that the resistivity saturation is also
seen in the insulating system at low enough carrier temperatures. We
have completely ignored the low temperature resistivity saturation
problem in our analysis, assuming, perhaps somewhat uncritically, but
in agreement with all existing theoretical analyses in the subject,
that the resistivity saturation arises from temperature
saturation, i.e. $T_s \equiv T_{\rm min}$.
Within the screening theory, in fact, thermal suppression
of screening will be cut off at temperatures ($T<T_D$) lower than the
temperature scale of the disorder scattering (with $T_D \geq \hbar/k_B
\tau$), and we have
taken this disorder broadening effect approximately
into account in the results
presented in Figs. 2 and 3 of this paper.
Another comment in the context of 2D MIT experiments
is the issue of the so-called separatrix, where the critical density
$n_c$ separating the metallic and the 
insulating phase is claimed to
manifest a completely temperature independent resistivity 
sharply delineating the metallic phase ($d\rho/dT >0$) from the
insulating phase ($d\rho/dT <0$) and thus defining a quantum phase
transition. Such a sharp separatrix cannot be explained at all within
our Drude-Boltzmann theory since all electronic states are by
definition extended in our semiclassical theory. As has been
emphasized above, the apparent critical density $n_c$ in our theory is
strongly temperature dependent with $n_c$ delineating
$d\rho/dT >0$ (for $n>n_c$) and $d\rho/dT <0$ (for $n<n_c$)
decreasing with the lowest measurement
temperature $T_{\rm min}$ with $n_c \rightarrow 0$ as $T_{\rm min}
\rightarrow 0$ since a true insulator does not exist in our theory. An
experimental separatrix defining the 2D MIT has, however, been rarely
reported, and in that sense our theory is consistent with a large body
of the existing 2D MIT experimental data. We do emphasize, however,
that at low enough densities all real 2D systems should eventually
exhibit exponential localization behavior (at densities below the
``power law'' apparent localization being discussed in this paper) which
is beyond the scope 
of our work. We believe that Si MOS systems, typically being more
disordered than 2D GaAs systems, exhibit much more of the
exponential-type true insulating behavior than the power-law apparent
insulating behavior being discussed in this paper.

In a very recent paper Noh {\it et al.} \cite{noh03} have 
reported an anomalous power law insulating behavior 
in a 2D p-GaAs systems at low densities (and intermediate
temperature), which we believe to be the apparent
insulating phase induced by quantum-classical crossover 
being predicted in this paper.
The $r_s$ values in the samples of ref. \onlinecite{noh03} are
extremely high ($r_s \sim 44 -80$), bringing up the possibility of
Wigner crystal (perhaps in a strongly correlated molten phase
\cite{spivak}) physics
playing a role in this experiment. 
We believe that it is unlikely that 
Wigner crystal (WC) physics has much to do
with this anomalous insulating behavior. First, 
the experimental temperature and density regime where the
anomalous insulating behavior
is observed is well
outside the WC phase boundary (see our Fig. 4 in this paper). 
Second, the transport data plotted as
$\sigma(T)\equiv \rho^{-1}$ for various densities in ref. \onlinecite{noh03}
show absolutely no indication of a
critical density separating metallic and insulating phases --- the
data at all densities look essentially the same at ``high
temperatures'' where $\sigma(T) \sim T$ whereas at ``low
temperatures'' the relatively high density conductivity shows an
upward bending below some density-dependent temperature $T_{\rm cross}(n)$
(i.e. $d\rho/dT$ changes sign from being insulating-like,  $d\rho/dT
<0$, at higher temperatures to being metallic-like, $d\rho/dT > 0$, at
lower temperatures). We emphasize that for $T>T_{\rm cross}(n)$ there is
absolutely no difference in the observed qualitative behavior of
$\sigma(T)$ at different densities. 
Since $T_{\rm cross}(n)$ decreases with carrier density $n$,
it is impossible to rule out the possibility that 
the experimental low
density $\sigma(T)$ plots would actually also bend upward at
temperatures lower than the lowest hole temperature ($\sim 60 mK$)
achieved in these experiments. In fact the experimental data of Noh
{\it et al.} are entirely consistent with this scenario. For example,
if the experimental temperature cut-off is taken to be
$200mK$ 
in the data of Noh {\it et al.}, then all the plots shown in
ref. \onlinecite{noh03} would appear to be ``insulating'' (upto the hole
density of $5\times 10^9 cm^{-2}$) since all the experimental data
exhibit $d\rho/dT <0$ (i.e. $d\sigma/dT >0$) down to 200mK.

To emphasize the similarity between the experimental data of Noh {\it
et al.} and our predicted crossover apparent insulating phase we
show in Fig. 2 our calculated low density transport results for the
Noh {\it et al}. sample. Remembering that phonon scattering
(neglected in our theory) becomes important for GaAs holes already
at 500 mK \cite{noh02}, the results of Fig. 2 are remarkably similar to the
experimental data presented in ref. \onlinecite{noh03}. This is
particularly true for Fig. 2(b), where we include a Dingle

\begin{figure}
\epsfysize=2.5in
\centerline{\epsffile{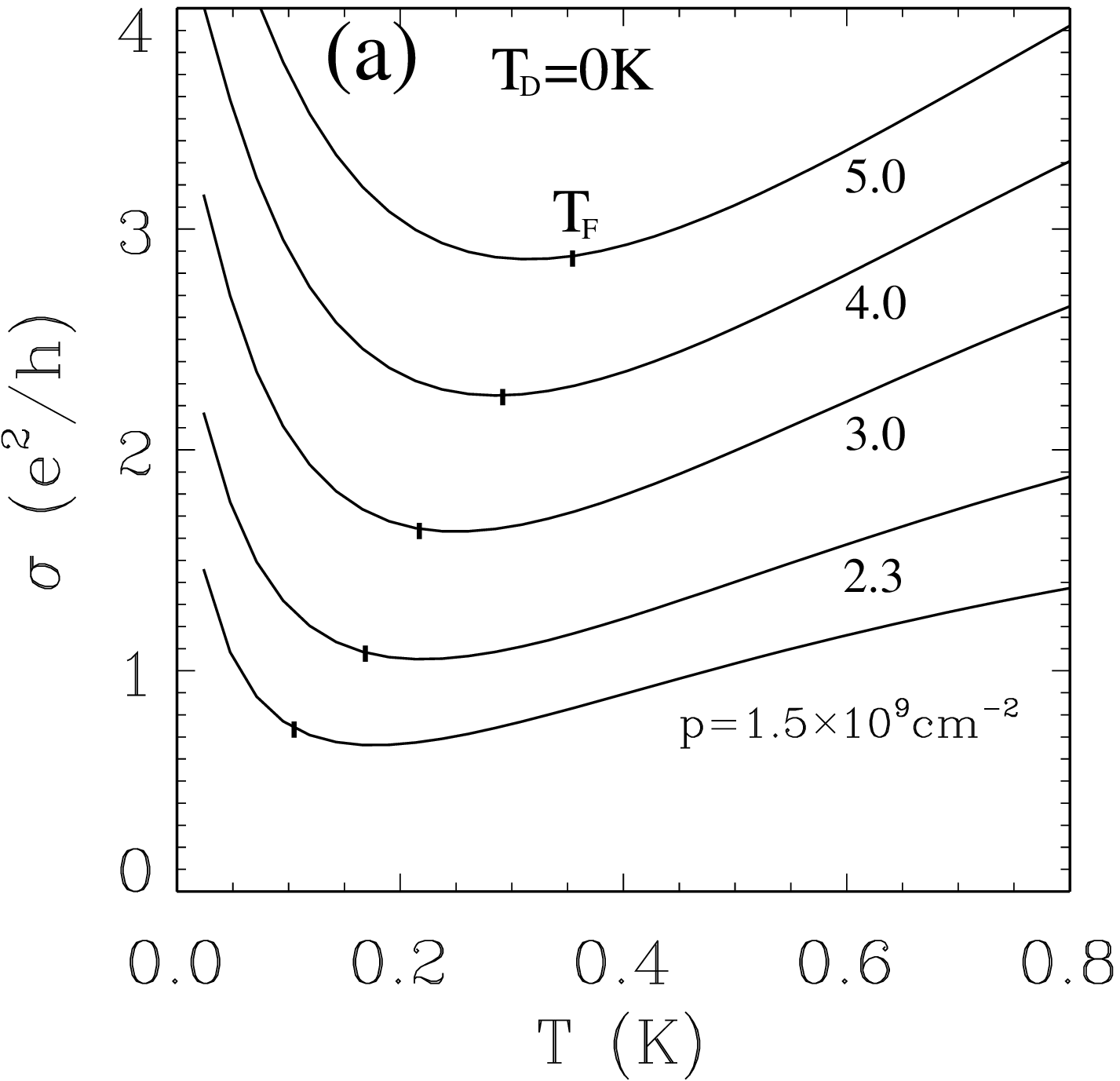}}
\epsfysize=2.5in
\centerline{\epsffile{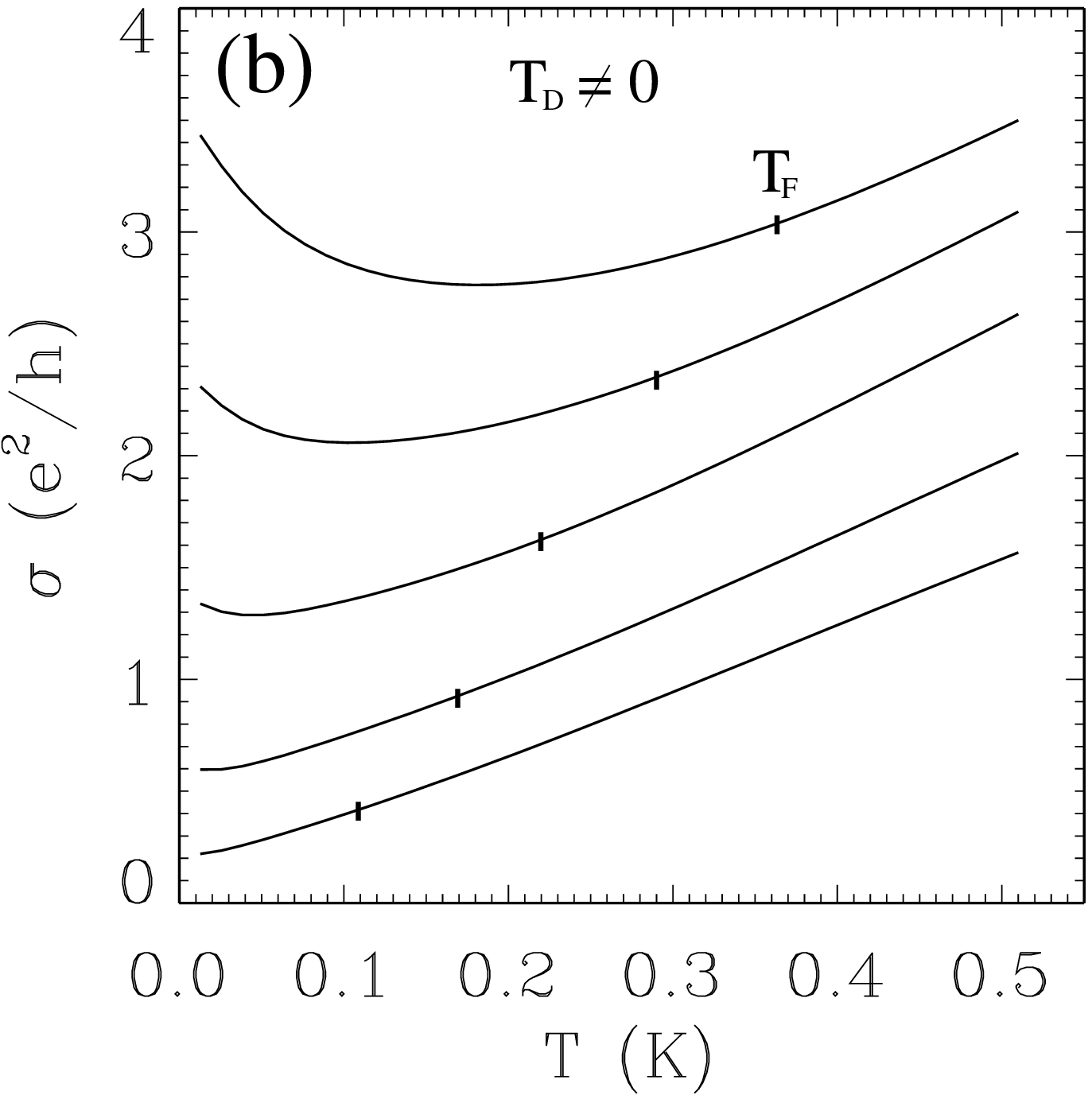}}
\caption{
Calculated conductivities
for p-GaAs system [7] for various hole
densities $p=$5.0, 4.0, 3.0, 2.3, 1.5$\times 10^{9}cm^{-2}$ (from the
top) with the remote charged impurities.
In (a) we calculate conductivities without level broadening in the
polarizability function, and in (b) we include the level broadening
in the screening, following ref. [14],
with $T_D=$0.2, 0.25, 0.3, 0.4, 0.5K (from the top).
The vertical bars indicate the Fermi
temperature $T_F$.}
\label{fig2}
\end{figure}

\noindent
temperature cut-off in the screening function in order to 
approximately simulate
the effect of collisional broadening (due to impurity scattering). The
basic idea\cite{DS1986} is to include in the finite wave vector static
polarizability the level broadening (parametrized by a Dingle
temperature $T_D = \Gamma/k_B$ where $\Gamma$ is the collisional
damping) arising from the impurity scattering. As discussed earlier in
the literature \cite{DS1986} such a collisional damping term,
parametrized by the level broadening parameter $T_D$, acts to suppress
the thermal effects on screening at low temperatures, $T\ll T_D$,
while leaving the screening function essentially unaffected at high
temperatures $T>T_D$. Since the level broadening parameter in the
screening, $T_D$, is unknown (within our approximation scheme), we do
not attach any great quantitative significance to our approximation
except to note that this zeroth order scheme of incorporating
collisional damping in screening is physically meaningful since
impurity scattering should tend to suppress finite wave vector static
screening of an electron gas. 
We emphasize that the parameter $T_D$ used in our theory should not be
construed to be $T_D \approx \hbar/k_B\tau$ (which would be far too
small), but should be considered an effective level broadening
parameter which cuts off the strong temperature dependence of
screening for $T<T_D$, with $T_D$ being a parameter of the theory
(increasing with decreasing conductivity).

It is therefore worth while to emphasize that our results (with $T_D
\ne 0$ in screening) shown in Fig. 2(b) are qualitatively (and even
semiquantitatively) in good agreement with the approximately 
linearly rising,
$\sigma \propto T$, conductivity at higher temperatures found by Noh
{\it et al}. In fact, even 
our $T_D=0$ static RPA screening transport
results are in qualitative agreement with experiment except for
the fact that the crossover temperature $T_{\rm cross}$ for $d\sigma/dT$
to change sign is consistently much higher in Fig. 2(a) with $T_D=0$
than in our Fig. 2(b) with $T_D \ne 0$ or in the experiment. The basic
behavior of an approximately linear temperature dependence of
conductivity, $\sigma \propto T$, for $T>T_{\rm cross}(n)$ with
$T_{\rm cross}(n)$ decreasing with decreasing density (and $d\sigma/dT
<0$, i.e. $\sigma(T)$ increasing with decreasing temperature for
$T<T_{\rm cross}(n)$) applies to Figs. 2(a) and (b) and to the
experimental data. Inclusion of phonon scattering (neglected in our
calculations here) is likely to further improve the good agreement
between experiment and theory.

In the context of comparison with the experimental data we show in
Fig. 3 our calculated low temperature transport results for the 2D
GaAs electron system corresponding to the recent measurement of Lilly
{\it et al.}\cite{lilly} carried out in a high quality low density
n-GaAs gated heterostructure system. We have kept the level broadening
parameter $T_D = 0.75K$ fixed in the results shown in Fig. 3 and have
shown the behavior of both the resistivity $\rho(T)$ and conductivity
$\sigma(T)$ for several low densities. Again, phonon scattering
effects, possibly of some 
importance here for $T>1K$ \cite{lilly}, have been
neglected. These theoretical results are again in striking 
qualitative
and semiquantitative agreement with the experimental data of Lilly
{\it et al}. \cite{lilly} demonstrating that the experimentally
observed anomalous insulating phase in high-quality
(i.e. low-disorder) 2D semiconductor systems 
may very well 
be the apparent crossover insulating phase (which would
manifest metallic behavior for $T<T_{\rm cross}(n)$ except that
$T_{\rm cross}$ may be too low to be experimentally accessible at low
carrier densities) predicted by our theory.

The striking qualitative agreement
between our theory and experiment for the anomalous 
insulating phase encourages us to predict the following
scenario for high quality GaAs-based 2D (both electron and hole)
systems. As 2D carrier density is lowered the high temperature
crossover insulating phase ($T>T_{\rm cross}$) extends to lower
temperatures as $T_{\rm cross}(n)$ goes down, and eventually when
$T_{\rm cross}$ goes below the lowest accessible experimental
temperature, the system appears to be insulating with an anomalous
(i.e. power law) temperature

\begin{figure}
\epsfysize=4.0in
\centerline{\epsffile{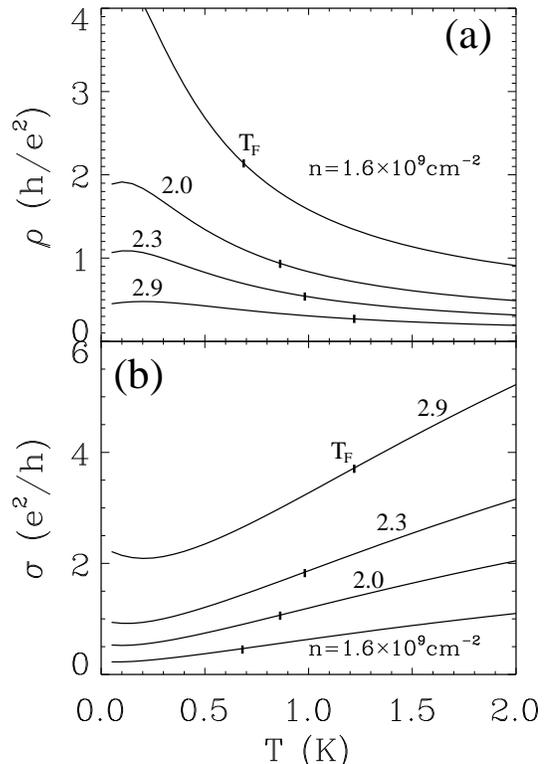}}
\caption{
Calculated (a) resistivities and (b) conductivities 
for n-GaAs system for various electron
densities $n=$1.6, 2.0, 2.3, 2.9$\times 10^{9}cm^{-2}$
with the remote charged impurities and a fixed Dingle temperature
$T_D=0.75K$
as a function of temperature. The vertical bars indicate the Fermi
temperature $T_F$.}
\label{fig3}
\end{figure}

\noindent
dependence characterizing the
apparent insulating phase. 
Since our predicted scenario is in excellent qualitative 
agreement with the available experimental data
in GaAs-based high-mobility 2D holes and electrons, we propose the
following direct experimental test for its verification (or
falsification): One should lower the temperature to check whether the
observed anomalous temperature dependence of
the putative insulating phase remains insulating or bends upward
(i.e. $d\sigma/dT$ goes through zero changing its sign from positive
to negative as it does at higher densities) at lower
temperatures. Since lowering carrier temperature  arbitrarily is
typically difficult to achieve in 2D semiconductor systems, a
relatively easy way of testing our proposed scenario will be to
directly plot the experimentally measured $d\sigma/dT$ (or $d\rho/dT$)
as a function of temperature at low densities in the anomalous
insulating phase. If the magnitude of this derivative is decreasing
with decreasing temperature indicating that $\sigma$ (or $\rho$) may
be approaching a minimum (or maximum), 
then this is a rather strong verification of
our proposed scenario. We mention in this context that the
experimental data of ref. \onlinecite{noh03} and \onlinecite{lilly}
are qualitatively consistent with our proposed scenario of a
decreasing magnitude of $d\sigma/dT$ as $T$ decreases.
We believe that there are simply two possibilities (which are
consistent with our prediction of the anomalous power law insulating
phase being a crossover phase): Either the magnitude of $d\rho/dT$
will decrease indicating a transition to an effective metallic phase
or $d\rho/dT$ will diverge exponentially indicating a true insulating
phase. We suggest that experiments be carried out in the anomalous
power law insulating phase (with careful measurements of $d\rho/dT$ as
a function of $T$ and $n$) to verify our prediction.

Finally, we discuss the possibility of WC physics playing a role in
the observed anomalous insulating behavior of refs.
\onlinecite{noh03,lilly}. In Fig. 4 we show our calculated approximate WC
($T-n$)
phase diagram for both 2D electrons and holes in GaAs, combining both
the $T=0$ quantum Wigner crystallization at low carrier densities and
the classical Wigner crystallization. The calculation of this
approximate WC phase diagram follows our recent work in
ref. \onlinecite{hwang} where we show how one can implement an
interpolation scheme to obtain the full phase diagram in the
density-temperature plot by combining the known classical\cite{r14} 
and quantum \cite{tanatar}
WC limits. We also show 
as shaded regimes the experimental samples
corresponding to refs. \onlinecite{noh03} and \onlinecite{lilly} in
Fig. 3(a) and (b) respectively. Both the shaded regions are well
outside the WC phase in the phase diagram indicating that WC physics
is unlikely to be playing a dominant role here. The more important
reason for ruling out the WC scenario is, however, experimental ---
the experimental data of Noh {\it at al.} \cite{noh03} is completely
smooth and the conductivity behavior in the ``metallic'' (higher
density) and the ``insulating'' (lower density) phase is identical at
higher temperatures, the only difference being the upward curvature in
$\sigma(T)$ in the higher density data which moves to lower
temperatures in the lower density plots and may have simply moved to
temperatures below the measurement temperatures at the lowest densities
making it experimentally invisible.

In summary, we have identified an apparent ``high temperature''
insulating phase in 2D systems at intermediate carrier densities where
the conductivity increases with 
temperature in a power law fashion
(approximately 
linearly). This apparent or effective insulating phase
is a temperature-induced quantum-classical crossover phase which
exists only for $T>T_{\rm cross}(n)$ where $T_{\rm cross}$ decreases with
decreasing carrier density. We have shown that, although typically
$T_{\rm cross} \sim T_F$ ($\propto n$), it is possible for 
$T_{\rm cross}$ to
be substantially below the Fermi temperature if collisional broadening
effects are included in the carrier screening function. We have
shown that our calculated transport behavior in this effective
insulating phase agrees qualitatively very well with recent
observations in high-quality 2D electrons and holes in GaAs
heterostructures, We have not discussed the 2D Si-MOS system in this
paper because the disorder effects in Si MOSFETs are substantially
higher (mobilities are typically factors of $10-1000$ lower in Si MOS
systems than in GaAs systems), and the Si MOS systems make transitions
to the strongly localized phase (with exponential temperature
dependence) much sooner --- this intermediate crossover insulating
phase therefore exists only in a very narrow

\begin{figure}
\epsfysize=2.7in
\centerline{\epsffile{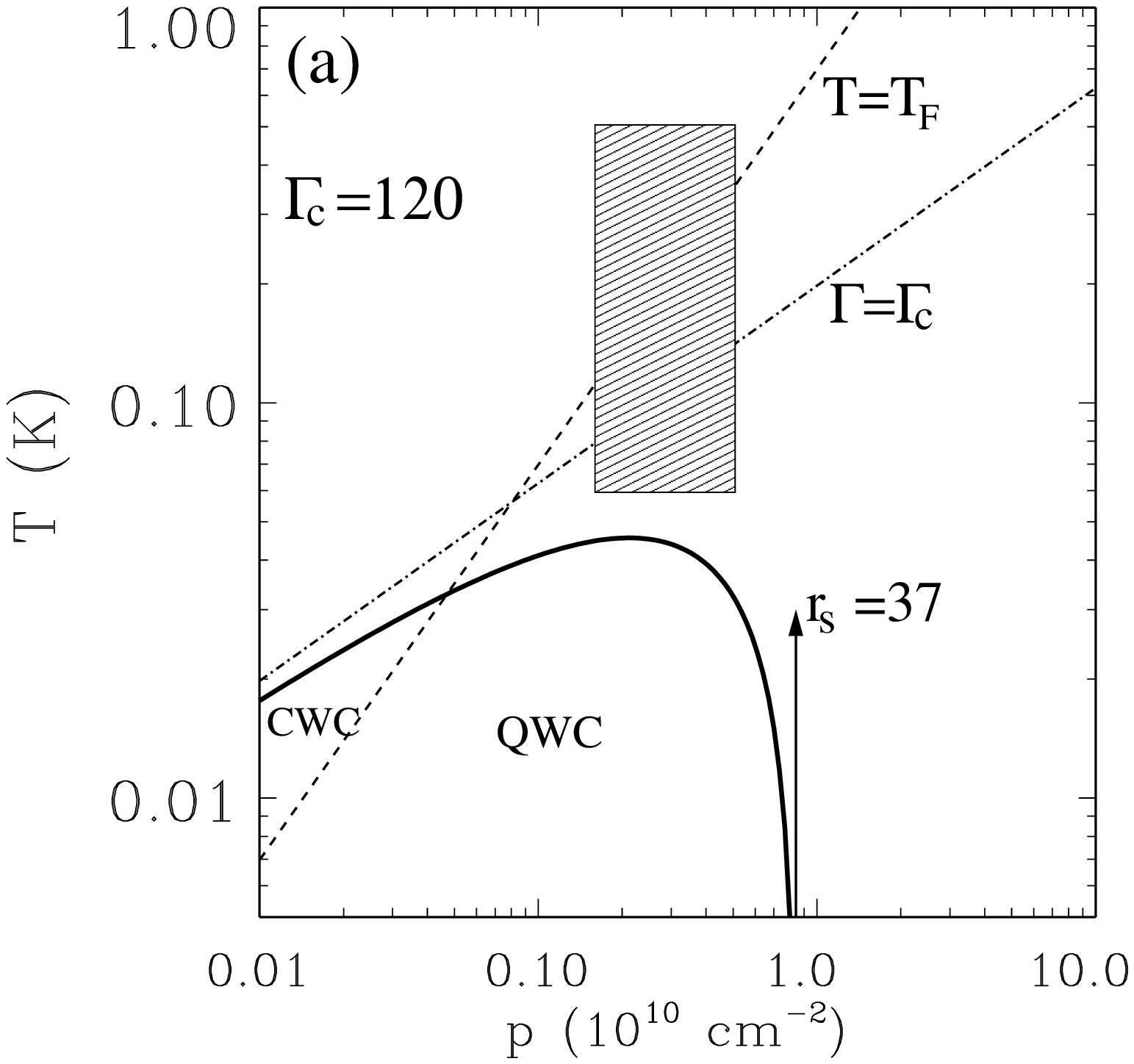}}
\epsfysize=2.9in
\centerline{\epsffile{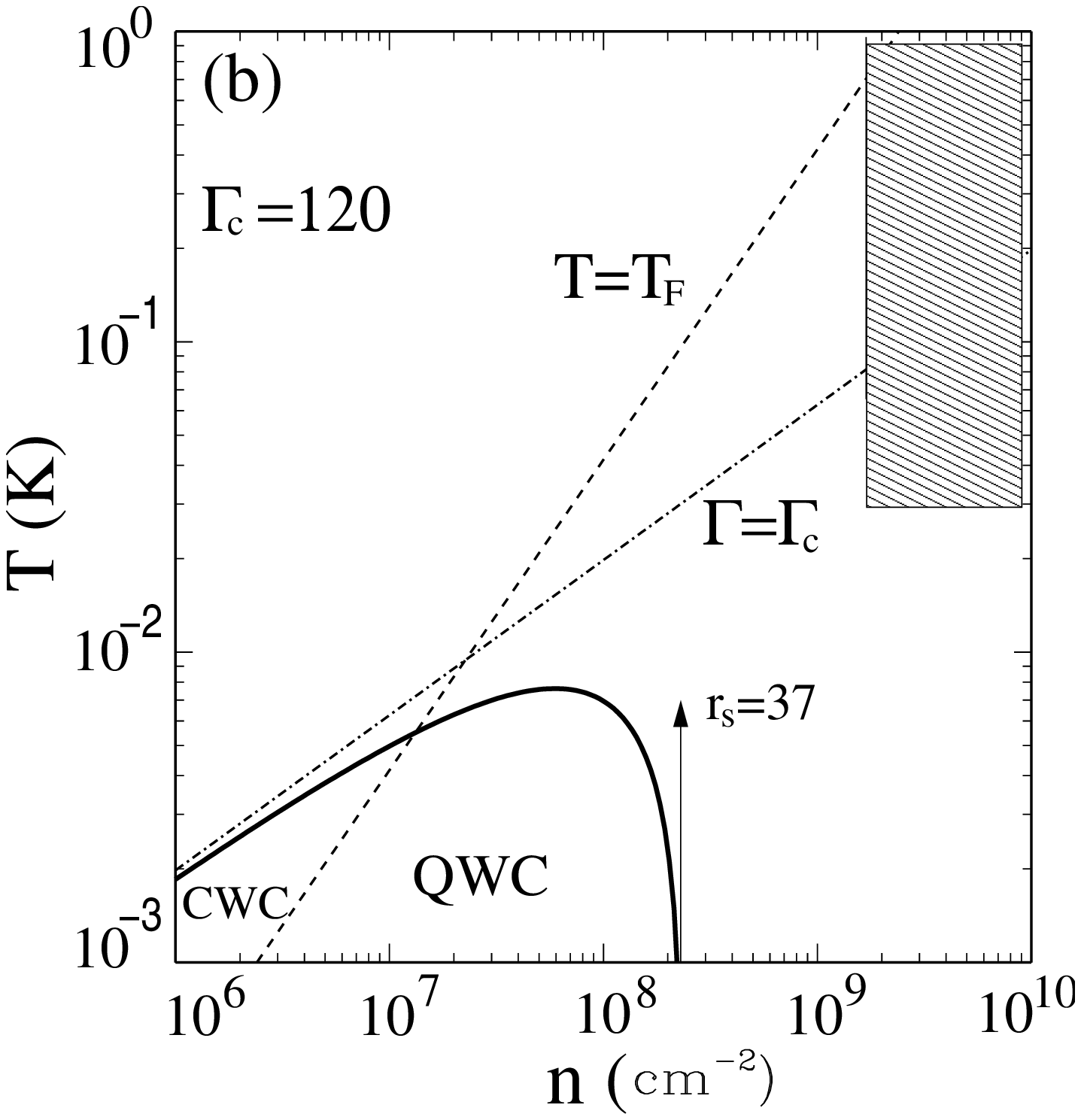}}
\caption{
Calculated 2D hole/electron liquid - Wigner crystal 
phase diagram in the density - temperature ($n,T$) plot for (a) p-GaAs
and (b) for n-GaAs systems.
The shaded regions indicate the experimental samples
corresponding to ref. [7] for p-GaAs and ref. [8] for n-GaAs.
Here $T_F$ is the Fermi temperature and 
$\Gamma =\langle V \rangle/\langle T \rangle$
is the ratio of the classical
mean potential energy to the mean classical kinetic energy.
The classical 
WC - electron liquid (first order) phase transition line is defined by 
$\Gamma = \Gamma_c$, which is found from numerical (molecular dynamics)
simulations [17] to be 
$\Gamma_c \approx 120$. 
At $T=0$ a 2D system is expected to become a
quantum WC phase at $r_s \ge 37$ [18].
The heavy solid line, indicating the liquid-solid phase boundary, is
obtained from the interpolating scheme described in ref. [16]. The
region above (below) the $T=T_F$ line is the classical (quantum)
region with CWC (QWC) denoting the classical (quantum) Wigner crystal
phase. 
}
\label{fig4}
\end{figure}

\noindent
range of densities and
temperatures in Si MOS systems.

The weakest empirical point in our quantitative comparison
with the experimental data is the fact that our theoretically obtained
$T_{\rm cross}^{th}(n)$ seems to be consistently higher than the
experimental crossover temperature 
$T_{\rm cross}^{ex}(n)$ where
$T_{\rm cross}$ at a particular density could for example be defined by
the condition $d\sigma/dT =0$ at $T=T_{\rm cross}$. We have tentatively
fixed this problem in this work by introducing the collisional damping
induced screening cut-off through the Dingle temperature parameter
$T_D$. The other possibility is to take the {\it real} Fermi
temperature $T_F$ to be smaller than the nominal $T_F$ ($\sim 
n/m $) given
by the carrier density $n$ and the band effective mass $m$ which could
happen if the actual effective mass is larger than the band mass
and/or the actual free carrier density is smaller (e.g. due to
trapping of carriers). While in the GaAs-based high mobility systems
studied in this paper, the discrepancy between $T_{\rm cross}^{th}(n)$ and
$T_{\rm cross}^{ex}(n)$ is relatively small even for the $T_D=0$ theory,
this discrepancy is very large in Si MOS based 2D systems where 
disorder is high. A straightforward application of our $T_D=0$ theory
for Si MOS systems indicates $T_{\rm cross} > 1K$ for the 
$n=5\times 10^{10}-10^{11} cm^{-2}$ density range of relevance to Si
2D systems. To reduce $T_{\rm cross}$ substantially one needs to introduce
large values of $T_D$ where the theory is not meaningful. The other
possibility is that the effective mass in Si may be enhanced from its
band value and/or the effective carrier density may be substantially
lower than the nominal $n$. It is therefore interesting to note that
there is no clear cut signature of this anomalous insulating phase in
2D Si MOS systems where the insulating phase, for $T<1K$ or so,
indicates exponentially rising resistivity with decreasing
temperature. This is consistent with our finding of rather large
$T_{\rm cross}^{th}$ in Si MOS systems implying that this apparent
insulating phase in Si systems remains only a high temperature phase
since the electrons can always be cooled below $T_{\rm cross}$ in the 2D
Si samples.

We note finally that the validity of
the semiclassical Drude-Boltzmann theory for studying the anomalous
insulating phase is not a crucial issue in this context. The reason is
that Boltzmann theory, being semiclassical, becomes increasingly valid
at higher temperatures, and therefore, should be a reasonable
qualitative description for the ``high-temperature'' effective
insulating phase being discussed here. The precise value of the
resistivity $\rho$ and the corresponding localization parameter $k_Fl$
(where $l$ is the transport mean-free path), i.e. whether $\rho$ is
smaller than the quantum of resistance $h/e^2$ or $k_F l$ is larger
than unity, are important issues for $T=0$ considerations (where
quantum localization effects are important) but not for finite $T/T_F$
($\approx 0.1-1$) case which is of interest here. In fact,
within our semiclassical Boltzmann transport theory the behavior shown
in Figs. 1 - 3 are {\it generic} behavior determined entirely by $n$
and $T$, completely independent of the actual resistivity values of
the 2D system, i.e. one should think of the ordinate ($\rho$ or
$\sigma$ in our figures) to have completely arbitrary units --- the
actual values of $\rho$ ($=\sigma^{-1}$) in our theoretical results
being determined by the unknown charged impurity density ($N_i$) in
the system which we use as an adjustable parameter to set the
resistivity scale. It is more appropriate to think of ordinate at
Figs. 1 and 2 as $\rho/\rho_0$ or $\sigma/\sigma_0$ where $\rho_0$ or
$\sigma_0$ are the $T=0$ Drude values of resistivity or
conductivity. These considerations make it clear that our qualitative
results are valid (at finite $T/T_F$) irrespective of the actual
sample resistivities as long as one can neglect quantum localization
effects which seems to be the case for the GaAs experiments in
refs. \onlinecite{noh03} and \onlinecite{lilly}, but perhaps not for Si
MOS samples.

In our conclusion we discuss critically the approximations of our
theory which may limit its applicability to real 2D systems. Our most
important approximations are the Drude-Boltzmann scattering theory
(due to charged impurity scattering) and RPA screening by the electron
liquid. Both of these approximations are simplistic at the low carrier
density phenomena of interest here. 
But there are good reasons to believe that our theory is
sound in explaining the qualitative behavior of 2D semiconductor
systems at densities and temperatures above the true localization
regime. First, recent systematic diagrammatic calculations \cite{zala}
of higher
order interaction corrections to the 2D resistivity show that the
basic picture of an effective 2D metallic behavior with $d\rho/dT >0$
(with a leading order linear temperature coefficient of resistivity)
applies well in the ballistic transport regime ($\hbar /k_B \tau \ll T
\ll T_F$)
provided weak localization effects are negligible. Second, the results
for $\rho(T,n)$ obtained in the 2D metallic regime within the
RPA-Drude-Boltzmann theory agree well with existing experimental
results. Third, RPA becomes quantitatively accurate at high
temperatures (and, our inclusion of local field correlation
corrections going beyond RPA give qualitatively very similar results).
We therefore believe that our description of the high-temperature
apparent insulating phase based on the semi-classical
Drude-Boltzmann-RPA transport theory may have considerable theoretical
validity.


We would like to thank  M. P. Lilly for helpful discussions.
This work is supported by the US-ONR and NSF-ECS.

\end{document}